\documentclass[12pt]{article}
\usepackage{amsmath}
\usepackage{amssymb}
\usepackage{dsfont}
\usepackage{cite}
\usepackage{enumerate}
\newsymbol \blackbox 1004
\newcommand{\eh}{\hfill}\newlength{\sperr}

\newenvironment{proof}{{\settowidth{\sperr}{\bf\rm
Proof}%
\par\addvspace{0.3cm}\noindent\parbox[t]{1.3\sperr}
{\bf\rm P\eh r\eh o\eh o\eh f.\eh }%
}}{\nopagebreak\mbox{}
$\blackbox$\par\addvspace{0.3cm}}

\def\a{\alpha}

\def\g{\gamma}

\def\s{\sigma}
\def\la{\lambda}

\def\vp{\varphi}

\def\ve{\varepsilon}
\def\wh{\widehat}
\def\wt{\widetilde}
\def\ov{\overline}

\def\BC{{\mathbb C}}
\def\BR{{\mathbb R}}
\def\BN{{\mathbb N}}

\def\cla{{\mathcal A}}

\def\clc{{\mathcal C}}

\def\clv{{\mathcal V}}

\def\col{\mathrm{col}}
\newcommand{\E}{\mathrm{e}}
\newcommand{\I}{\mathrm{i}}

\newtheorem{Pa}{Paper}[section]
\newtheorem{Tm}[Pa]{{\bf Theorem}}

\newtheorem{Cy}[Pa]{{\bf Corollary}}
\newtheorem{Rk}[Pa]{{\bf Remark}}

\newtheorem{Ee}[Pa]{{\bf Example}}
\newtheorem{Dn}[Pa]{{\bf Definition}}

\newtheorem{Pn}[Pa]{{\bf Proposition}}

\title{Generalized wave operators: dynamical  and stationary cases and divergence problem}

\author{Lev Sakhnovich\footnote{99 Cove ave. Milford, CT, 06461, USA. E-mail: lsakhnovich@gmail.com}}

\date{}
\begin{document}
\maketitle

 \begin{abstract} 
 Ideas and results of the generalized wave
operator theory for dynamical  and stationary cases are developed further  and 
exact expressions for generalized scattering operators are obtained for wide classes of 
differential equations. New results on the structure of the generalized scattering operators
are derived. Interesting interrelations
between dynamical  and stationary cases are found for  radial Schr\"odinger and Dirac equations,
and for Dirac-type equations as well.
 For some important examples we explain why the well-known divergences 
 in the higher order approximations of the scattering matrices do not appear in 
 the "generalized wave operator"  approach.
  \end{abstract}
	
  

\section{Introduction}\label{Intro}
Wave  and scattering operators belong to the basic notions of mathematical physics and spectral theory.
Three books studying this topic (in the mentioned above domains)  \cite{AlK, EH, Str} as well as hundreds of papers were published in 2015 only.
{\it Generalized wave operators} were introduced in  the 1960s \cite{Do, Sakh8}, whereas interesting nontrivial examples appeared already in \cite{Do}
and much wider classes were considered in \cite{Sakh8}.  The theory of generalized wave operators was actively developed at the end of the 1960s
 and at the beginning of the 1970s 
(see, e.g., \cite{Ber, BuM, JMG, Mat, OC, Pear, Sakh2}). Using generalized wave operators one can consider scattering theory for wider classes of important operators
and in greater detail than using the classical  wave operators.  More precisely, in many important cases the initial and final states of the system cannot
 be regarded as free (i.e., no free states at $t=\pm{\infty}$). In these cases the theory of   generalized wave
and {\it generalized scattering operators} proves to be effective.

Several interesting papers (see, e.g. \cite{Tong, Xi}) as well as an important
book by  L.D. Faddeev and S.P. Merkuriev \cite{FaMe} on generalized wave operators appeared in the 1980s. However, numerous
open problems remain and various further important results and applications may be obtained. In the recent years, the ideas developed in the
theory of  generalized wave operators are starting to attract attention again (see \cite{GiVe, KuKu, Nak, Wir} and references therein).
In particular, some ideas are used in the theory of modified wave operators (the case of nonlinear differential equations).
At the same time various other results and publications on the spectral and scattering theory of radial Dirac and Schr\"odinger equations
and equations with singularities appeared last years (see, e.g., \cite{AlK, BEKT, KoSaTe,  SaSaR} and references therein).
In the present paper we plan to demonstrate new important applications of the generalized wave operators to scattering theory.

Ideas and results of the generalized wave
operator theory for dynamical  and stationary cases are developed further  and 
exact expressions for scattering operators are obtained for wide classes of 
differential equations. New results on the structure of the generalized scattering operators
are derived. Interesting interrelations
between dynamical  and stationary cases are found for  radial Schr\"odinger and Dirac equations,
and for Dirac-type equations as well.
 A "generalized wave
operator"  approach to the well-known divergences
 in the higher order approximations of the scattering matrices is discussed,
 and for some important examples we explain why these divergences
 do not appear in  our approach.

In Section \ref{Prel} we introduce the notions of   the generalized wave
and generalized scattering operators and of the {\it deviation factors}
 (see \cite{Ber}, \cite{JMG}, \cite{Sakh8} and \cite{Sakh2}).  The deviation factors
describe the deviation of the initial and final states from
the free state. In Section \ref{Coul} we consider the case of Schr\"odinger equation with Coulomb potential.

Section \ref{Erg} is dedicated to the radial Schr\"odinger equation.
The generalized scattering matrix $S_{dyn}$ describes the behavior of a system when
 the time tends to infinity (dynamical case). In the present paper we introduce also the
generalized scattering matrix $S_{st}$ which describes the behavior of a system when the space coordinate tends to infinity (stationary case). We discuss the connections between $S_{dyn}$ and
$S_{st}$ and derive the corresponding ergodic-type theorem in Section \ref{Erg}.
Similar results for the radial Dirac equation are obtained in Section \ref{radDir}
(and for Dirac-type equation in Section \ref{Dity}).

Dirac equation in $\BR^4$ is studied in Section \ref{S6}. Under some natural conditions we prove that the  scattering matrix of the 
Dirac operator in $\BR^4$ (momentum representation) 
has a special structure (see \eqref{5.19}) which was not known before. This structure follows from the fact that   the generalized scattering matrix commutes with the unperturbed operator.
We note that the unperturbed Dirac operator (in the momentum representation) is the operator of multiplication by the matrix function $H(q)$
given by \eqref{9.9}. 

In the last part of the paper (Sections \ref{S7}-\ref{Exp}), we investigate the fundamental equation of quantum electrodynamics. 
The structure of the scattering operator for  the fundamental equation of quantum electrodynamics
is given in Theorem \ref{Theorem 6.1}.
In the electrodynamic theory, the higher order approximations of
matrix elements of the scattering matrix contain integrals which diverge.
We think that these divergences appear because the  corresponding scattering matrix is introduced
in the form of the small parameter series.
We produce exact formulas  for the scattering
matrices and  for some important  examples we prove (see Section \ref{Coul} and, especially, Sections  \ref{S8} and \ref{Exp})
that the divergences do not exist in the case of "generalized wave
operator"  approach.
In our approach we essentially use the  results  from Section \ref{Prel}.

The last section of the paper is Conclusion.

As usual, $\BR$ stands for the real axis and $\BC$ stands for the complex plane, $\overline{\lambda}$ stands for the complex number which
is complex conjugate to $\lambda$, $I$ stands for the identity operator, $\E$
 is the base of the natural logarithm, $\I$ is the imaginary unit and Span stands for linear span.

\section{Preliminaries: generalized wave operators} \label{Prel}
\setcounter{equation}{0}
Consider linear (not necessarily bounded)
operators $A$ and $A_{0}$ acting in some Hilbert space $H$  and assume that the operator $A_{0}$
is self-adjoint.  The absolutely continuous subspace of the operator $A_0$ (i.e., the subspace corresponding
to the absolutely continuous spectrum) is denoted by $G_0$, and $P_0$ is the orthogonal projector on   $G_0$.
Generalized wave operators $W_{+}(A,A_{0})$ and $W_{-}(A,A_{0})$ are introduced by the equality
\begin{equation}W_{\pm}(A,A_{0})
=\lim_{t{\to}\pm\infty}[\E^{\I At}\E^{-\I A_{0}t}W_{0}(t)^{-1}]P_0,
\label{1.1}\end{equation}
where $\I$ is the imaginary unit and $W_0$ is an operator function taking operator values $W_0(t)$   acting in $G_0$
in the domain $|t|>R$ ($t \in \BR$) for  some $R \geq 0$.  
More precisely, we have the following definition (see \cite{Sakh2, Sakh8}) of
 the generalized wave operators $W_{\pm}(A,A_{0})$ and deviation factor $W_0$.
 \begin{Dn}\label{Definition 1.1}
An operator function $W_{0}(t)$ is called a deviation factor 
and operators $W_{\pm}(A,A_{0})$ are called generalized wave operators if
the
following conditions are fulfilled:
\begin{enumerate}[1.]
	\item The operators $W_{0}(t)$ and  $W_{0}(t)^{-1}$  acting in  $G_0$,
are
bounded for all $t$ \, $(|t|>R)$, and
\begin{equation}\lim_{t{\to}\pm\infty}W_{0}(t+\tau)W_{0}(t)^{-1}P_0=P_0,\quad\tau=\overline{\tau}.
\label{1.2}
\end{equation}
\item The following commutation relations  hold for arbitrary values $t$ and $\tau$:
\begin{equation}W_{0}(t)A_{0}P_0=A_{0}W_{0}(t)P_0,\quad
W_{0}(t)W_{0}(t+\tau)P_0=W_{0}(t+\tau)W_{0}(t)P_0.
\label{1.3}\end{equation}
\item The limits $W_{\pm}(A,A_{0})$ in \eqref{1.1}
exist in the sense of strong convergence. 
\end{enumerate}
\end{Dn}
If  $W_{0}(t) \equiv I$ in $G_0$, then the operators $W_{\pm}(A,A_{0})$ are usual wave
operators.
\begin{Dn}\label{Definition 1.7} 
The operators $A$ and $A_0$ are called comparable in the generalized sense if
the generalized wave operators $W_{\pm}(A,A_0)$ and $W_{\pm}(A_0,A)$ exist.
\end{Dn}
Although the notion of  the generalized wave operator was introduced  in \cite{Do, Ber, JMG, Sakh8},
its description in the form of Definition \ref{Definition 1.1} was given several years later  in \cite{BuM, Sakh2}.
\begin{Pn}\label{Proposition 1.2} 
Let conditions \eqref{1.1}--\eqref{1.3} be
fulfilled. Then
\begin{equation} W_{\pm}(A,A_{0})\E^{\I A_{0}t}P_0=\E^{\I At} W_{\pm}(A,A_{0})P_0
\label{1.4}\end{equation}\end{Pn}

\begin{proof}
Rewrite \eqref{1.1} in the form
\begin{equation}W_{\pm}(A,A_{0})=\lim_{t{\to}\pm\infty}[\E^{\I A(t+s)}\E^{-\I A_{0}(t+s)}W_{0}(t+s)^{-1}P_0].
\label{1.5}\end{equation}
Now, formula \eqref{1.2},  the second equality in \eqref{1.3} and equality \eqref{1.5}
 imply that
 \begin{equation}W_{\pm}(A,A_{0})P_0=\E^{\I As}W_{\pm}(A,A_{0})\E^{-\I A_{0}s}P_0.
\label{1.6}\end{equation}
This proves the proposition.
\end{proof}

\begin{Dn}\label{Definition 1.4} Let conditions \eqref{1.1}--\eqref{1.3} be
fulfilled.
Then the generalized wave operators $W_{\pm}(A,A_0)$ are called complete if
\begin{equation}W_{\pm}(A,A_0)G_0=G_A,\label{1.8}\end{equation}
where $G_A$ is the absolutely continuous subspace of the operator $A$.
\end{Dn}

Using \eqref{1.6}, we obtain the following assertion (see \cite[Ch. 7]{AG}).
\begin{Pn}\label{Proposition 1.5}Let conditions \eqref{1.1}--\eqref{1.3} be
fulfilled.
If the operator $W_{+}(A,A_0)$ is complete, then
\begin{equation}A_{a}W_{+}(A,A_0)f=W_{+}(A,A_0)A_{0,a}f,\quad
f{\in}G_0,\label{1.9}\end{equation}
where $A_a$ and $A_{0,a}$ are the operators induced by $A$ and $A_0$ in the spaces
$G_A$ and $G_0$, respectively.
\end{Pn}

\begin{Tm}\label{Theorem 2.6} 
Let  self-adjoint operators  $A_0$, $A_1$ and $A$ be given. If the operator  $A_1-A$ belongs
to the
trace class $\sigma_1$ and the
generalized wave operators
\begin{equation}
W_{\pm}(A,A_{0})
=\lim_{t{\to}\pm\infty}[\E^{\I At}\E^{-\I A_{0}t}W_{0}(t)^{-1}]P_0
\label{1.10}\end{equation}
exist and are complete, then the  generalized wave
operators
\begin{equation}W_{\pm}(A_1,A_{0})
=\lim_{t{\to}\pm\infty}[\E^{\I A_{1}t}\E^{-\I A_{0}t}W_{0}(t)^{-1}]P_0
\label{1.11}\end{equation}
exist and are complete as well.
\end{Tm}

\begin{proof} 
We need the well-known Rosenblum--Kato theorem  \cite{Ros, Kat}:

\noindent
\emph{If the  operators $A$ and $A_1$ are self-adjoint and the operator  $A_1-A$ belongs
to the
trace class $\sigma_1$, then the wave operators
\begin{equation}W_{\pm}(A_1,A) =\lim_{t{\to}\pm\infty}[\E^{\I A_{1}t}\E^{-\I At}]P_A,
\label{1.12 }\end{equation}
where $P_A$ is the orthogonal projector on $G_A$,
exist and are complete.}\\
Using Rosenblum--Kato approach and theorem, it is easy to show that the operators $W_{\pm}(A,A_{0})$ map the subspace
 $G_0$ onto $G_A$. Hence,
\begin{equation}
W_{\pm}(A_1,A_0)=W_{\pm}(A_1,A)W_{\pm}(A,A_0).
\label{1.13}\end{equation}
The theorem is proved.
\end{proof}

Clearly, the choice of the deviation factor is not unique.

\begin{Rk}\label{Remark 1.8} 
Let unitary operators $C_{-}$ and $C_{+}$ satisfy commutation conditions $A_{0}C_{\pm}=C_{\pm}A_{0}$. If $W_{0}(t)$ is a deviation factor, then
the operator function given (for $t>0$ and $t<0$, respectively) by the equalities
$W_{+}(t)=C_{+}W_{0}(t)$\, ($t>0$), and $W_{-}(t)=C_{-}W_{0}(t)$\, ($t<0$) is the deviation factor as well.
\end{Rk}
The choice of the operators $C_{\pm}$ is very important and is determined by specific physical problems.
The definition below shows that generalized scattering operators also depend on the choice of $C_{\pm}$.
\begin{Dn}\label{Definition 1.9}
The generalized scattering operator $S(A,A_0)$ has the form
\begin{equation}S(A,A_0)=W^{*}_{+}(A,A_0)W_{-}(A,A_0),\label{1.14}
\end{equation}
where
\begin{equation}W_{\pm}(A,A_{0})
=\lim_{t{\to}\pm\infty}\Big( \E^{\I At}\E^{-\I A_{0}t}W_{\pm}(t)^{-1}\Big) P_0 .
\label{1.15}\end{equation}
\end{Dn}

In fact, operator functions $W_{\pm}(t)$ are uniquely determined up to some factors $C_{\pm}(t)$
tending to $C_{\pm}$ when $t$ tends to $\infty$ or $- \infty$, respectively. This means that
 $S(A,A_0)$ is uniquely determined by the choice of $C_{\pm}$.
 
It is not difficult to prove that the operator $S(A,A_0)$ unitarily  maps $G_0$  onto itself and that
\begin{equation}
A_{0}S(A,A_0)P_{0}=S(A,A_0)A_{0}P_{0}.
\label{1.16}
\end{equation}

\section{Coulomb potential}\label{Coul}
\setcounter{equation}{0}
\paragraph{1.} 
We introduce the operators
\begin{equation}
\mathcal{L}_{1}f=-\frac{d^{2}}{dx^{2}}f+\left( \frac{\ell(\ell+1)}{r^{2}}-
\frac{2z}{r}\right) f,\quad
\mathcal{L}_{0}f=-\frac{d^{2}}{dx^{2}}+\frac{\ell(\ell+1)}{r^{2}}f,
\label{2.1}
\end{equation}
where $z>0$, $\ell$ is some nonnegative integer (i.e., $\ell \in \BN_0$), and
the  boundary condition is given by
\begin{equation} f(0)=0.\label{2.2}\end{equation}
In {\it momentum representation},  $\mathcal{L}_{0}$ and $\mathcal{L}_{1}$ have the following form
 \cite[Sec. 8]{BetSal}:
\begin{equation} \widetilde{\mathcal{L}}_{0}f=k^{2}f(k),\quad\quad \qquad f{\in}L^{2}(0,\infty),\label{2.3}
\end{equation}
\begin{equation} \widetilde{\mathcal{L}}_{1}f=k^{2}f(k)+\int_{0}^{\infty}f(p)R_{\ell}(k,p)dp.\label{2.4}
\end{equation}
Here $R_{\ell}(k,p)$ is given by the equality
\begin{equation}R_{\ell}(k,p)=-\frac{2z}{\pi}Q_{\ell}\left(\frac{k^{2}+p^{2}}{2pk}\right) ,
\label{2.5}\end{equation}
and $Q_{\ell}(x)$ is Legendre function of the second kind. The deviation  factors in the momentum representations are operators of multiplication
by functions of the form (see \cite{Sakh2}):
\begin{equation}\widetilde W_{\pm}(t,k)=|t|^{\mp(\I z/k)},\label{2.6}\end{equation}
which we sometimes call deviation factors as well.
Thus, the generalized scattering operator exists and can be written in the form
\begin{equation}S(\widetilde{\mathcal{L}}_{1},\widetilde{\mathcal{L}}_{0})=\lim |t|^{-\I z/k}S(t,\tau)|\tau|^{-\I z/k},\quad\tau{\to}-\infty,\quad t{\to}+\infty ,
\label{2.7}\end{equation}
where
\begin{equation}S(t,\tau)=
\E^{\I t\widetilde{\mathcal{L}}_{0}}\E^{-\I t\widetilde{\mathcal{L}}_{1}}
\E^{\I \tau\widetilde{\mathcal{L}}_{1}}\E^{-\I \tau\widetilde{\mathcal{L}}_{0}}.\label{2.8}\end{equation}
We have shown in \cite{Sakh2} that the scattering operator $S(\widetilde{\mathcal{L}}_{1},\widetilde{\mathcal{L}}_{0})$ has the form
\begin{equation}
S(\widetilde{\mathcal{L}}_{1},\widetilde{\mathcal{L}}_{0})f(k)=S(k)f(k),\quad k>0,\label{2.9}
\end{equation}
where  $S(k)$ is introduced by the equality
\begin{equation}S(k)=(2k)^{4 \I z/k} \, \, \frac{\Gamma(\ell+1-\I z/k)}{\Gamma(\ell+1+\I z/k)}.
\label{2.10}
\end{equation}
Here $\Gamma(x)$ is the Euler gamma function.

Formulas \eqref{2.9} and  \eqref{2.10} present an expression for the dynamical scattering
operator. On the other hand, we note that the stationary scattering problem is very well known
for the case of Coulomb potential \cite{Bur}. In particular, for the stationary
scattering operator $S_{st}$ the following equality holds \cite[Ch. 1]{Bur}:
\begin{equation}S_{st}(k)=\frac{\Gamma(\ell+1-\I z/k)}{\Gamma(\ell+1+\I z/k)}.
\label{2.11}\end{equation}
The  difference between the functions $S(k)$ and $S_{st}(k)$ consists only in the  factor $(2k)^{4 \I z/k}$,
which does not depend on $\ell$. Such a difference is insignificant and is caused by the choice of normalization. 
This fact confirms that the introduced definitions are correct
from the physical point of view.
In Section~\ref{Erg} and further we discuss some other important stationary
scattering problems  (and in greater detail).

\paragraph{2.}  Our further results in this section are useful for understanding interconnections between removal
of divergence problem and generalized scattering matrix.
Expanding the right-hand side of \eqref{2.7} in powers of $z$  we obtain the normalized perturbation series
\begin{equation}S(\mathcal{ \wt L}_{1}, \mathcal{\wt L}_{0})=I+\sum_{i=1}^{\infty}z^{i}
\widetilde{S}_{k}(+\infty,-\infty) .
\label{2.12}
\end{equation}
The first term of the normalized series is defined by the relation
\begin{align} \label{2.13}
\Big(\widetilde{S}_{1}(+\infty,-\infty)f\Big)(k)=& \lim_{\substack{\tau{\to}-\infty \\
t{\to}+\infty}}
\Bigg( -\big( \I/k\big) \Big(\ln\big| t\tau\big|\Big) f(k)
\\ & \nonumber
+\frac{2\I}{\pi}
\int_{\tau}^{t}\int_{0}^{\infty}f(p)Q_{\ell}\left( \frac{k^{2}+p^{2}}{2pk}\right) 
\E^{\I \big( k^{2}-p^{2}\big)t_{1}}dpdt_{1}\Bigg) .
\end{align}
It follows from \eqref{2.10} that
\begin{equation}\Big(\widetilde{S}_{1}(+\infty,-\infty)f\Big)(k)=-\frac{2 \I z}{k}
\left(
\frac{\Gamma^{\prime}(\ell+1)}
{\Gamma(\ell+1)}-2\ln{(2k)}\right)f(k).
\label{2.14}
\end{equation}

\paragraph{3.}
 Now, let us expand  (in powers of $z$)  the right-hand side of \eqref{2.9}:
\begin{equation}S(t,\tau)=I+\sum_{i=1}^{\infty}z^{i}
{S}_{k}(t,\tau).\label{2.15}\end{equation}
It is easy to see that
\begin{equation}S_{1}(t,\tau)f=\frac{2\I}{\pi}
\int_{\tau}^{t}\int_{0}^{\infty}f(p)Q_{\ell}\left(\frac{k^{2}+p^{2}}{2pk}\right)
\E^{\I (k^{2}-p^{2})t_{1}}dpdt_{1}.
\label{2.16}
\end{equation}
Relations  \eqref{2.13}, \eqref{2.14} and \eqref{2.16} imply that
\begin{equation}
S_{1}(t,\tau)f=\Big( 
\big(\I /k\big) \big(\ln{|t\tau|}\big) +O(1)\Big) f(k),\,k>0,\,t{\to}+\infty,
\,\tau{\to}-\infty.\label{2.17}\end{equation}
\begin{Rk}\label{Remark 2.1}
The operator $S_{1}(t,\tau)$ has a logarithmic type divergence. Using
generalized scattering operator we obtain the results  $($see \eqref{2.10} and \eqref{2.14}$)$ without divergences.
\end{Rk}

\section{Classical and generalized stationary scattering problems (Schr{\"o}dinger equation)}\label{Erg}
\setcounter{equation}{0}
\paragraph{1  Classical and Coulomb cases.}
Stationary scattering results are well-known in the classical and Coulomb cases \cite{Bur}.
The classical radial Schr{\"o}dinger equation has the form
\begin{equation}
\left(\frac{d^2}{dr^2}-\frac{\ell(\ell+1)}{r^2}+U(r)+k^2\right) y_{\ell}(r)=0  \qquad \big(\, U(r)=\overline{U(r)} \, \big),
\label{3.1}
\end{equation}
where
\begin{equation}
 \int_{0}^{\infty}r^{p}|U(r)|dr<\infty \quad (p=1,2).\label{3.2}
\end{equation}
There exists a solution of \eqref{3.1} which satisfies the following
boundary conditions
\begin{align}& y_{\ell}(r){\sim}Nr^{\ell+1},\quad r{\to}0,\label{3.3}
\\ &
y_{\ell}(r){\sim}\exp{(-\I {\theta}_{\ell})}-\exp{(\I {\theta}_{\ell})}S_{\ell}(k),\quad r{\to}\infty,\label{3.4}\end{align}
where $N$ is a constant,  ${\theta}_{\ell}=kr-\frac{1}{2}\ell\pi$ and
$S_{\ell}(k)$ is \emph{the scattering function}.

The radial Schr{\"o}dinger equation with Coulomb potential has the form
\begin{equation}
\left( \frac{d^2}{dr^2}-\frac{\ell(\ell +1)}{r^2}+U_{c}(r)+k^2\right) y_{\ell}(r)=0,
\label{3.5}\end{equation}
where
\begin{equation}U_{c}(r)=\frac{2z}{r},\quad z>0.\label{3.6}\end{equation}
Again, there exists (see \cite[Ch. 1]{Bur}) a solution of \eqref{3.5} satisfying
the boundary conditions 
\begin{align}& y_{\ell}(r){\sim}Nr^{\ell +1},\quad r{\to}0,\label{3.7} \\
& y_{\ell}(r){\sim}\exp{(-\I {\theta}_{\ell}^{c})}-\exp{(\I{\theta}_{\ell}^{c})}
S_{\ell}^{c}(k),\quad r{\to}\infty,\label{3.8}\end{align}
where $N$ is a constant and  ${\theta}_{\ell}^{c}=kr-\frac{1}{2}\ell\pi-(z/k)\ln(2kr)$.
The Coulomb stationary scattering function  was already considered in Section \ref{Coul}
and is given by the formula \eqref{2.11}. Below we consider a more general case.

\paragraph{2 Generalized stationary scattering function.}
Let us consider the radial Schr{\"o}dinger equation
\begin{equation}
\left( \frac{d^2}{dr^2}-\frac{\ell(\ell+1)}{r^2}+\varphi(r)+k^2\right) y_{\ell}(r)=0 \qquad \big(\varphi(r)=\overline{\varphi(r)}\,\big),
\label{3.10}\end{equation}
where
\begin{equation}\int_{a}^{\infty}|\varphi^{\prime}(r)|dr+
\int_{a}^{\infty}|\varphi^{2}(r)|dr+ \int_{0}^{a}|\varphi(r)|rdr<\infty,\quad 0<a<\infty.\label{3.11}\end{equation}
It is easy to see that the Coulomb potential satisfies  \eqref{3.11}.
In the present (more general) case, 
 a solution of \eqref{3.10} satisfying
the boundary conditions \eqref{3.7} exists as well.
We use the following result \cite[Ch. II, Theorem 8]{Bel}:\\
 \emph{Under conditions \eqref{3.11},  equation \eqref{3.10} has two linear independent solutions}
\begin{align}&
Z_{1}(r,k.\ell){\sim}\exp{(-\I {\theta})}V_{0}^{-1}(r,k),\quad r{\to}\infty , 
\label{3.14}
\\ &
Z_{2}(r,k.\ell){\sim}\exp{(\I{\theta})}V_{0}(r,k),\quad r{\to}\infty , 
\label{3.15}\end{align}
\emph{where} $\,\, \theta=kr-\frac{1}{2}\,\ell\pi \,\,$ \emph{and}
\begin{equation}V_{0}(r,k)=\exp{\left( \frac{\I}{2k}\int_{a}^{r}\varphi(u)du\right) }.
\label{3.16}\end{equation}
 Hence,   we obtain  our main assertion in this section.
\begin{Pn}\label{Proposition 3.1}Let \eqref{3.11} be fulfilled. Then the solution $y_{\ell}$ of \eqref{3.10}, which
satisfies \eqref{3.7}, has the form 
\begin{equation}y_{\ell}(r){\sim}\exp{(-\I {\theta})}V_{0}(r,k)^{-1}-\exp{(\I {\theta})}V_{0}(r,k)S_{\ell}(k),
\quad r{\to}\infty,\label{3.17}\end{equation}
where   $V_{0}(r,k)$ is
the deviation factor given by \eqref{3.15}.\end{Pn}
\begin{Dn}\label{StCase} The functions $S_{\ell}(k)$ and $V_{0}(r,k)$ in \eqref{3.17}
are  called the stationary generalized  scattering function and deviation factor $($in the case of radial Schr\"odinger equation$)$, respectively.
\end{Dn}

\begin{Rk}\label{Remark 3.2}The deviation factor $V_{0}(r,k)$
characterizes the deviation of the wave $y_{\ell}(r)$ from the free wave.
It is important that $V_{0}(r,k)$ does not depend on $\ell$.
\end{Rk}

\paragraph{3.}
 Let us consider the operators
\begin{equation}
Hf=\left( -\frac{d^2}{dr^2}+\frac{\ell(\ell+1)}{r^2}-\varphi(r)\right) f,\quad f(0)=0
\label{3.18}\end{equation}
and
\begin{equation}H_{0}f=-\frac{d^2}{dr^2}f,\quad f(0)=0.
\label{3.16!}\end{equation}
According to Buslaev--Matveev  results \cite{BuM}, the following assertion is valid.

\begin{Pn}\label{Proposition 3.3}
Let the potential $\varphi(x)$ and its derivatives satisfy the conditions
\begin{equation}
\big| \varphi^{(\kappa)}(x)\big| {\leq}C(1+x)^{-\alpha-\kappa},\quad \alpha>1/2,\quad  0{\leq}\kappa{\leq}2.
\label{3.19}
\end{equation}
Then there exists the generalized wave operator $W_{+}(H,H_0)$, and the corresponding deviation factor $($in momentum representation$)$ is given by the formula
\begin{equation}\widetilde W_{0}(t,k)=\exp{\left( \frac{\I}{2k}\int_{a}^{tk}\varphi(u)du\right)} .
\label{3.20}
\end{equation}
\end{Pn}

Thus, we derived that the stationary deviation factor $V_0(r)$, when $r{\to}\infty$, and the dynamical deviation factor,
 when $t{\to}\infty$, are connected by the important {\it ergodic} equality
 \begin{equation}
V_0(tk,k)=\widetilde W_0(t,k).
\label{3.21}\end{equation}
The following remark is an anologue (for the stationary case) of Remark \ref{Remark 1.8}.
\begin{Rk}\label{Remark 3.4} Let a unitary operator $C$  be such that $A_{0}C=CA_{0}$.
If $V_{0}(r)$ is a deviation factor, then
$V(r)=CV_{0}(r)$ is a deviation factor as well.\\
The choice of the operator $C$ depends on the specific physical problem.\end{Rk}

\section{Generalized stationary scattering problems (radial Dirac and Dirac-type systems)}\label{radDir}
\setcounter{equation}{0}
\paragraph{1.}
Radial Dirac system has the form
\begin{align}&
\left( \frac{d}{dr}+\frac{k}{r}\right) f-\big(\lambda+m-v(r)\big) g=0,
\label{4.1}
\\ &
\left( \frac{d}{dr}-\frac{k}{r}\right) g+\big(\lambda-m-v(r)\big) f=0 \quad (k>0,\quad m>0),
\label{4.2}\end{align}
where $\lambda = \ov{\la}$ and
\begin{equation}
v(r)=-\frac{A}{r}+\varphi(r)\qquad (A>0,\quad |k|>A, \quad \vp(r)=\ov{\vp(r)}\, ).
\label{4.3}\end{equation}
We assume that the following inequality
\begin{equation}
\int_{a}^{\infty}|\varphi^{\prime}(r)|dr+
\int_{a}^{\infty}|\varphi^{2}(r)|dr+ \int_{0}^{a}|\varphi(r)|dr<\infty,\quad 0<a<\infty\label{4.4}\end{equation}
is fulfilled. Then, there exists (see \cite{Sakh11}) a solution of \eqref{4.1}, \eqref{4.2} satisfying
the boundary condition 
\begin{equation}\col [f,g]{\sim}r^{\alpha}\col [1,b_0],\quad r{\to}0,
\label{4.5}\end{equation}
where col stands for column and
\begin{equation}
\alpha=\sqrt{k^2-A^2},\quad b_0=(\alpha+k)/A,\quad \Re{\alpha}>0.
\label{4.6}\end{equation}
Let us rewrite the system \eqref{4.1}, \eqref{4.2}  in the matrix form:
\begin{equation}
\frac{dZ}{dr}=\cla(r)Z,\quad \cla(r)=\left[\begin{array}{cc}
                  -k/r & m+\lambda-v(r)\\
               m-\lambda+v(r)    & k/r
                \end{array}\right] , \quad  Z=\begin{bmatrix} f \\ g \end{bmatrix}.
\label{4.7}\end{equation}
We use  the following result (see \cite[Ch. II, Theorem 8]{Bel}).\\
 \emph{The system \eqref{4.7} has two linearly independent solutions}:
\begin{align}&
Z_{1}(r,\lambda,k){\sim}\exp{(-\I {\theta})}V_{0}(r,\lambda)^{-1}C_{1}(\lambda,k),\quad r{\to}\infty , \label{4.9}
\\ &
Z_{2}(r,\lambda,k){\sim}\exp{(\I {\theta})}V_{0}(r,\lambda)C_{2}(\lambda,k),\quad\quad\, r{\to}\infty , \label{4.10}\end{align}
\emph{where ${\theta}={\eta}r$, \,$\eta=\sqrt{\lambda^{2}-m^{2}}$,\, $|\lambda|>m,$
 $C_1(\lambda,k)$ and $C_2(\lambda,k)$ are $2{\times}1$ vectors,  $C_1(\lambda,k)=\overline{C_2(\lambda,k)}$,  and}
\begin{equation}V_{0}(r,\lambda)=\exp{\left( \I \frac{\lambda}{\eta}\int_{a}^{r}v(u)du\right) } .
\label{4.11}
\end{equation}
 Hence,   we obtain  the following assertion.

\begin{Pn}\label{Proposition 4.1} Let  \eqref{4.4} hold and let  $\col [{f,g}]$ be the solution of \eqref{4.7}
satisfying \eqref{4.5}. Then, for $r \to \infty$, we have
\begin{equation}\col [{f,g}]{\sim}\exp{(-\I {\theta})}V_{0}(r,\lambda)^{-1}C_1(\lambda,k)+\exp{(\I{\theta})}V_{0}(r,\lambda)C_2(\lambda,k)
,\label{4.12}\end{equation}
where  ${\theta}={\eta}r$, \, $|\lambda|>m.$
\end{Pn}
Consider the entries of  $C_1(\lambda,k)=\col [c_{11}(\lambda,k),c_{21}(\lambda,k)].$
Taking into account  \eqref{4.1}, \eqref{4.2} and \eqref{4.9}  we have
$\frac{c_{11}}{c_{21}}=\I\sqrt{\frac{\lambda+m}{\lambda-m}}$,
and so
\begin{align}&\label{d1}
c_{11}\big/\overline{c_{11}}=-c_{21}\big/\overline{c_{21}} \quad (|\la | >m).
\end{align}
\begin{Dn}\label {Definition 4.2}  The scattering matrix
 function $S(\lambda, k)$ of system \eqref{4.1}, \eqref{4.2} $($where \eqref{4.4} holds$)$
 is defined by the relation
 \begin{equation}S(\lambda,k)=
  \begin{bmatrix}-\overline{c_{11}(\lambda,k)}\big/c_{11}(\lambda,k) & 0 \\
 0 & \overline{c_{11}(\lambda,k)}\big/ c_{11}(\lambda,k) 
 \end{bmatrix} \label{4.13}
 \end{equation}\end{Dn}
Recall that $C_1(\lambda,k)=\overline{C_2(\lambda,k)}$. Hence, equalities \eqref{d1} and \eqref{4.13}  yield
\begin{equation}S(\lambda,k)C_1(\lambda,k)=-C_2(\lambda,k).\label{4.14}
\end{equation}
In view of \eqref{4.14}, equality  \eqref{4.12} (for $|\la | >m$ and  $r{\to}\infty$) can be rewritten in the form
\begin{equation}\col [{f,g}]{\sim}
\Big(\exp{(-\I {\theta})}V_{0}(r,\lambda)^{-1}-\exp{(\I{\theta})}V_{0}(r,\lambda)S(\lambda,k)\Big) C_1(\lambda,k)
  .\label{4.15}\end{equation}

\begin{Rk}\label{Remark 4.3} We emphasize that the deviation factor $V_{0}(r,\lambda)$ does not depend on $k$.
\end{Rk}
\begin{Rk}\label{Rk4.4}
Comparing  expression \eqref{4.11} for $V_{0}(r,\la)$ with  the corresponding formula
for $\widetilde W_{0}(t,p)$ from  \cite[Theorem 2.2]{Sakh12}, we obtain our next ergodic equality
\begin{equation}
V_0(ts, \la)= \widetilde W_0(t,p),\label{d3}\end{equation}
where $\la=\sqrt{p^2+m^2}$ and $s=p/ \lambda$.
\end{Rk}
\paragraph{2.}
Next, consider Dirac-type system:
\begin{align}&\left(\frac{d}{dr}+a(r)\right) f-\big(\lambda+m-b(r)\big) g=0,
\label{4.16}
\\ &
\left(\frac{d}{dr}-a(r)\right) g+\big(\lambda-m-b(r)\big) f=0 \quad (m>0), \label{4.17}\end{align}
where $a(r)=\ov{a(r)}$ and $b(r)=\ov{b(r)}$. We assume that $a$ and $b$ satisfy the inequalities
\begin{equation}
\int_{1}^{\infty}|b^{\prime}(r)|dr+
\int_{1}^{\infty}|b^{2}(r)|dr+ \int_{0}^{1}|b(r)|dr<\infty, \label{4.18}\end{equation}
\begin{equation}
\int_{a}^{\infty}|a^{\prime}(r)|dr+
\int_{1}^{\infty}|a^{2}(r)|dr+ \int_{0}^{\infty}|a(r)|dr<\infty .\label{4.19}\end{equation}
Then, there exists a solution $\col [f, \, g]$ of the system  \eqref{4.16}, \eqref{4.17} (written down in matrix form), which satisfies
the boundary condition
\begin{equation}\col [f(0),\, g(0)]=\col [0, \,1].\label{4.20}\end{equation}
If inequalities \eqref{4.18} and \eqref{4.19} hold, then there are  linearly independent solutions
$Z_1$ and $Z_2$ of the Dirac-type system, which admit representations \eqref{4.9} and \eqref{4.10}, respectively.  The corresponding deviation factor
$V_{0}(r,\lambda)$, in this case, has the form
\begin{equation}V_{0}(r,\lambda)=\exp{\left( \I\frac{\lambda}{\eta}\int_{1}^{r}b(u)du\right)}I_2 ,
\label{4.11!}\end{equation}
where (similar to the equality \eqref{4.11}) $\eta=\sqrt{\lambda^{2}-m^{2}}$.
The next proposition follows.
\begin{Pn}\label{Proposition 4.4}
Let  inequalities \eqref{4.18} and \eqref{4.19} hold.
 Then formulas \eqref{4.13} and \eqref{4.15} are valid for the Dirac-type case as well.
\end{Pn}
 \section{Generalized dynamical scattering problem (Dirac-type system)}\label{Dity}
\setcounter{equation}{0}
\paragraph{1.}  Introduce  the Dirac operator $\mathcal{L}$ acting in $L^{2}_{2}(0,\infty)$:
\begin{align}& \Big(\mathcal{L}\Psi\Big)(r)=\left(J\frac{d}{dr}+B(r)\right)\Psi(r),
\label{11.1}
\\&
J:=\left[\begin{array}{cc}
                    0 & -1 \\
                    1 & 0
                  \end{array}\right],\quad
B(r):=\left[\begin{array}{cc}
           m+b(r) & a(r) \\
           a(r) & -m+b(r)
         \end{array}\right].\label{11.2}
                  \end{align}
Then, we can rewrite Dirac-type system \eqref{4.16},      \eqref{4.17}        in the form
\begin{align}&
\label{d2}
\mathcal{L}\Psi =\la \Psi, \quad \Psi=\begin{bmatrix} f \\ g
\end{bmatrix}.
\end{align}
The operator $\mathcal{L}_{0}$ is introduced by the equality
\begin{equation}\big(\mathcal{L}_{0}\Psi\Big)(r)=\left(J\frac{d}{dr}+B_{0}(r)\right)\Psi(r), \quad
B_{0}:=\left[\begin{array}{cc}
           m & 0 \\
           0 & -m
         \end{array}\right].
\label{11.3}\end{equation}
The boundary condition for the operators $\mathcal{L}_{0}$ and $\mathcal{L}$ has  the form
\begin{equation}f(0)=0.\label{11.5}\end{equation}
Next, we determine $a$ and $b$ on the semiaxis $(-\infty, 0)$ by the equalities $a(-r)=a(r)$
and $b(-r)=b(r)$ ($r>0$), and consider operators $\mathcal{L}$ and $\mathcal{L}_{0}$
given by \eqref{11.1}, \eqref{11.2}  and by \eqref{11.3}, respectively, and acting in $L^{2}_{2}(-\infty,\infty)$. 
The operator $\mathcal{L}_{0}$
in the momentum representation has the form
\begin{equation}\big(\widetilde{\mathcal{L}}_{0}\widetilde{\Psi}\big)(p)=H_{0}(p)\widetilde{\Psi}(p), \quad H_{0}(p):=\left[\begin{array}{cc}
                               -m & -\I p \\
                               \I p & m
                             \end{array}\right],
\label{11.9}
\end{equation}
where $\widetilde{\Psi}(p){\in}L^{2}_{2}(-\infty,\infty)$.
The matrix $H_{0}(p)$ admits representation
\begin{equation}H_{0}(p)=U(p)D(p)U(p)^{-1},\label{11.11}\end{equation}
where
\begin{align}& U(p)=c\left[\begin{array}{cc}
                             \I s & 1\\
                             1 &   \I s
                           \end{array}\right],\quad
D(p)=\left[\begin{array}{cc}
             \mu & 0 \\
             0 & -\mu
           \end{array}\right],\label{11.12}
\\ &           
           \mu=\sqrt{p^2+m^2},\quad s=p/\mu,\quad c=(1+s^2)^{-1/2}.
\label{1.13!}\end{align}
\paragraph{2.}
Applying  to the radial Dirac equation the approach, which V.S. Buslaev and V.B. Matveev used
for the Schr\"odinger (see \cite{BuM}), we obtain the following important theorem.
\begin{Tm}\label{Theorem 11.1} Let  $a(r)$ and $b(r)$
 satisfy $($for $r>0$, fixed values $\a$ and $\g$, and for $\nu$ taking values $0,1$ and $2)$ the inequalities
\begin{align}&\big( |a^{(\nu)}(r)|+|b^{(\nu)}(r)|\big) {\leq}C(1+r)^{-\alpha-\nu},\quad 1>\alpha>3/4;
\label{11.6}
\\ &
|a(r)|<C(1+r)^{-1-\alpha+\gamma},\quad 1-\alpha<\gamma<\alpha - 1/2 .\label{11.7}\end{align}
Then the generalized wave operators $W_{\pm}(\mathcal{L},\mathcal{L}_{0})$ exist
and the corresponding deviation factor   $W_{0}(t)$ in the momentum representation
has the form
\begin{equation}\widetilde{W}_{0}(t,p)=\exp\left(\I \, {\mathrm{sgn}}(t) \int_{1}^{|t|}b(rs)dr\right)I_{2}. \label{11.8}\end{equation}
\end{Tm}
\begin{proof}
First  introduce the operator function
\begin{equation}\Theta(t)=\exp(\I t\mathcal{L})\exp(-\I t\mathcal{L}_{0})W_{0}(t).
\label{11.14}\end{equation}
In order to prove the  theorem, it  suffices to show that the  equality
\begin{equation}\left\|\frac{d\Theta(t)}{dt} \Psi\right\|=O\left(t^{1+\varepsilon}\right) \qquad (\varepsilon>0,
\quad t{\to}\infty) \label{11.15}\end{equation}
holds (for $t>0$) on some set  which is dense in  $L_{2}^{2}(-\infty,\infty)$.
Introduce the set $S$ of vector functions $\Psi\in L_{2}^{2}(-\infty,\infty)$ such that their Fourier transforms
(i.e., images in the momentum space)
\begin{equation}\widetilde{\Psi}(p)=\frac{1}{\sqrt{2\pi}}\int_{-\infty}^{\infty}
\E^{-\I pr}\Psi(r)dr\label{1.16!}\end{equation}
belong to the class $C^{\infty}$ and
\begin{equation}{\mathrm{supp}} (\widetilde\Psi) {\subset}\{ p:\,\, 0<c_1(\Psi)<|p|<c_2(\Psi)<\infty\}.
\label{11.17}\end{equation}
Further we consider $\Psi \in S$ of the forms $\Psi=\col [h, \, 0]$ and $\Psi=\col [0, \, h]$ with  Fourier transforms  $\wt \Psi=\col [\wt h, \, 0]$ and $\wt \Psi=\col [0, \, \wt h]$,
respectively.
Using momentum representation of $\frac{d\Theta}{dt}\Psi$ (for $\Psi$ mentioned above)
we obtain
\begin{equation}\left\|\left(\frac{d\Theta}{dt}\Psi\right)(r,t)\right\|{\leq}\|J_1(r,t)\|+\|J_2(r,t)\|,
\label{11.18}\end{equation}
where
\begin{align}& J_1(r,t)=\int_{-\infty}^{+\infty}(b(r)-b(st))
\exp\{\I tF(p,r,t)\}\widetilde{h}(p){dp},\label{11.19}
\\ &
J_2(r,t)=\int_{-\infty}^{+\infty}{a(r)}
\exp\{\I tF(p,r,t)\}\widetilde{h}(p){dp}.\label{11.20}\end{align}
Here $s$ is given in \eqref{1.13!} and
\begin{equation}F(p,r,t)=pr/t-\mu-(1/t)\int_{1}^{t}b(us)du.\label{11.0}\end{equation}
Integrating the integrals in \eqref{11.19} and \eqref{11.20} by parts, we obtain (in the same way it was done in \cite{BuM})  the inequalities
\begin{equation}\left(\int_{|r|{\leq}\varepsilon t}|J_{k}^{2}(r,t)|dr\right)^{1/2}{\leq} \, C_1 t^{-(\alpha+1/2)}
\qquad (k=1,2),\label{11.21}\end{equation}
which hold for some $C_1>0$. Now, let us estimate the function $J_1(r,t)$ in the domain $|r|{\geq}\varepsilon t.$ The stationary-phase point $p_{0}(r,t)$ is the solution of the equation
\begin{equation}\frac{\partial}{{\partial}p}F(p,r,t)=0.
\label{11.23}\end{equation}
Thus, we have
\begin{equation}r-ts-\frac{d \, s}{d \, p}\int_{1}^{t}ub^{\prime}(us)du=0 \qquad  \left(b^{\prime}(z)=\frac{d \, b}{d \, z}\right),
\label{11.24}\end{equation}
where $s(p)$ is given in \eqref{1.13!}. Using  \eqref{11.6}, we rewrite  \eqref{11.24} in the form
\begin{equation}r-ts+O(t^{1-\alpha})=0.
\label{11.25}\end{equation}
Hence, the stationary-phase point $p_{0}(r,t)$ is such that
\begin{equation}ts_{0}(r,t)=r+O(t^{1-\alpha}) \qquad \Big(s_0=p_0/\mu_0=p_{0}\Big/\sqrt{p_{0}^2+m^2}\Big).
\label{11.26}\end{equation}
Let us consider the case when $p>0$ and $t>0$. 
We say that $s$ belongs to the domain  $\Delta(r,t)$
if
\begin{equation}
|r-st| \, {\leq}\,  C_2 t^{\gamma} \label{11.28}
\end{equation}
for some fixed $\g$ satisfying \eqref{11.7} (and for some fixed $ C_2>0$).
In particular, the point  $s_0=p_{0}/\sqrt{p_{0}^2+m^2}$ belongs to   $\Delta(r,t)$ for all sufficiently large values of $t$. Using \eqref{11.6} and
the Langrange's mean value equality
\begin{equation}b(r)-b(st)=b^{\prime}(\xi)(r-st),\label{11.27}
\end{equation}
 we obtain
\begin{equation}|b(r)-b(st)|=O(t^{-1-\alpha+\gamma}),\quad s{\in}\Delta(r,t).\label{11.29}\end{equation}
When $s{\in}\Delta(r,t)$, relations \eqref{11.7} and \eqref{11.29} can be written in the form
\begin{equation}|a(r)|+|b(r)-b(st)|=O\Big(\big((r+1)t\big)^{-1/2-\alpha/2+\gamma/2}\Big),\quad s{\in}\Delta(r,t).\label{11.30}\end{equation}
Similar to \eqref{11.19} and \eqref{11.20}, we introduce the integrals $J_k$ on $\Delta$:
\begin{align}& J_1(\Delta)=\int_{\Delta}\big(b(r)-b(st)\big)
\exp\{\I tF(p,r,t)\}\widetilde{h}(p){dp},\label{11.31}
\\ &
J_2(\Delta)=\int_{\Delta}{a(r)}
\exp\{\I tF(p,r,t)\}\widetilde{h}(p){dp}.\label{11.32}\end{align}
It follows from \eqref{11.30} that
\begin{equation}\left(\int_{r{\geq}\varepsilon t}\left| J_{k}^{2}(\Delta(r,t))\right| dr\right)^{1/2}{\leq} \, C_3 t^{\gamma-\alpha-1/2}
\quad (k=1,2),\label{11.33}\end{equation}
where, according to \eqref{11.7}, we have
\begin{equation}\gamma-\alpha-1/2<-1.\label{11.34}\end{equation}
Now, consider domain $\widetilde{\Delta}(r,t)$  of values of $s$ such that
\begin{equation}
s\, {\notin} \, {\Delta}(r,t), \quad
|r|{\geq} \varepsilon t  .\label{11.35}\end{equation}
We write down $J_1\big(\widetilde{\Delta}(r,t)\big)$ in the form 
\begin{align} \label{11.35+} & J_1\big(\widetilde{\Delta}(r,t)\big)= J_{11}\big(\widetilde{\Delta}(r,t)\big)-J_{12}\big(\widetilde{\Delta}(r,t)\big),
\\ &
J_{11}\big(\widetilde{\Delta}(r,t)\big):=\int_{\widetilde{\Delta}(r,t)}b(r)
\exp\{\I tF(p,r,t)\}\widetilde{h}(p){dp},\label{11.36}
\\ &
J_{12}\big(\widetilde{\Delta}(r,t)\big):=\int_{\widetilde{\Delta}(r,t)}b(st)
\exp\{\I tF(p,r,t)\}\widetilde{h}(p){dp}.\label{11.37}\end{align}
Integrating  $J_{11}$ and $J_2$ by parts,  we obtain the inequalities
\begin{equation}|J_{11}\big(\widetilde{\Delta}(r,t)\big)|{\leq}|b(r)|t^{-1-\gamma},
\quad |J_{2}\big(\widetilde{\Delta}(r,t)\big)|{\leq}|a(r)|t^{-1}.
\label{11.38}\end{equation}
It follows from \eqref{11.6} and \eqref{11.38} that
\begin{equation}\left(\int_{r{\geq}\varepsilon t}\left| J_{11}^{2}\big(\widetilde{\Delta}(r,t)\big)\right|dr\right)^{1/2}+
\left(\int_{r{\geq}\varepsilon t}\left|J_{2}^{2}\big(\widetilde{\Delta}(r,t)\big)\right|dr\right)^{1/2}{\leq} \, C_4 t^{-1/2-\alpha}.
\label{11.39}\end{equation}
Next,   integrating by parts the expression $J_{12}$, we derive
\begin{equation}J_{12}\big(\widetilde{\Delta}(r,t)\big)=\frac{1}{t}\int_{\widetilde{\Delta}(r,t)}B(s,r,t)
\exp\{\I pr\}dp,\label{11.40}\end{equation}
where $B(s,r,t)=
\frac{\partial}{{\partial}p}\left(b(st)\widetilde{h}(p)\Big/ \frac{\partial}{{\partial}p}F(p,r,t)\right)$, and so
\begin{equation}|B(s,r,t)| {\leq}\, C_5 t^{-\alpha}(t/r).
\label{11.41}\end{equation}
Hence, we have
\begin{equation}\left(\int_{r{\geq}\varepsilon t}\left|J_{12}^{2}\big(\widetilde{\Delta}(r,t)\big)\right|dr\right)^{1/2}
{\leq} \, C_6 t^{-1/2-\alpha}.
\label{11.42}\end{equation}
Finally, by virtue of  \eqref{11.18}, \eqref{11.21}, \eqref{11.33}, \eqref{11.34},\eqref{11.39} and \eqref{11.42},
we easily obtain \eqref{11.15}. The theorem is proved.
 \end{proof}
Comparing \eqref{4.11!} and \eqref{11.8},  we  see that again (similar to the non-relativistic case)
 the stationary deviation factor $V_0(r,\lambda)$ and the dynamic deviation factor
 $\wt W_0(t,p)$ are connected by a simple equality
 \begin{equation}
 V_0(ts, \sqrt{p^2+m^2} \, )=\clc(p)\wt W_0(t,p) \quad (t>0),\label{11.43}\end{equation}
where $| \clc(p)|=1$.
Equality \eqref{11.43} shows that the class of  deviation factors for a
Dirac-type stationary problem
coincides with the  class of deviation factors for the corresponding dynamical problem. 

\section{Dirac equation in $\BR^4$}\label{S6}
\setcounter{equation}{0}
\paragraph{1.}
The classical  Dirac equation in the space of four variables has the form
\begin{equation}
\left( 
-e\varphi(x,t){I}_4+mc^2{\beta}+\sum_{n=1}^{3}{\alpha}_{n}(cp_{n}+e A_{n}(x,t))\right)
\psi(x,t)=\I \hbar
\frac{\partial\psi(x,t)}{\partial{t}},\label{9.1}\end{equation}
where $\psi(x,t)$ is the wave $4{\times}1$ vector function for the particle of the rest mass $m$, $c$ is the speed of light, 
$\hbar$
is the Planck constant divided by $2\pi$, $x \in \BR^3$ and $t$ are  the space-time coordinates, $p_k=-\I \hbar \frac{\partial}{\partial{x_k}}$, $e$ is the charge of the particle, $\varphi$ is a scalar potential, $\mathbf{A}=\col [A_1,A_2,A_3]$ is a vector potential. The Hermitian $4{\times}4$ matrices $\alpha_{k}$ and $\beta$ satisfy the relations
\begin{equation}
\alpha_{k}^{2}=\beta^{2}=I_{4} ; \quad \alpha_{k}\beta+\beta\alpha_{k}=0; \quad\alpha_{k}\alpha_{\ell}+\alpha_{\ell}\alpha_{k}=0 \quad (k{\ne}\ell).
\label{9.2}\end{equation}
Without loss of generality we put
\begin{align} &
 \beta=        \left[  \begin{array}{cc} I_2 & 0 \\
                       0 & -I_2
                     \end{array}\right], \quad 
\alpha_1=\left[  \begin{array}{cc}    0 & I_2 \\
                       I_2 & 0
                     \end{array}\right]; \quad \alpha_k=\left[  \begin{array}{cc}    0 & \s_k \\
                       \s_k & 0
                     \end{array}\right] \quad (k=2,3),
                     \label{9.5}
\\          &
             \s_2:=        \left[  \begin{array}{cc}
                       0 & -\I  \\
                       \I & 0
                     \end{array}\right],\quad  
\s_3:=          \left[  \begin{array}{cc}
                       1 & 0 \\
                       0 &-1 \end{array}\right] ,
                     \label{9.6}\end{align}
where $\s_k$ are Pauli matrices.                      
It is convenient to rewrite  \eqref{9.1} in the form
\begin{equation}
\left( mc^2{\beta}+c\sum_{n=1}^{3}{\alpha}_{n}p_{n}+V(x,t)\right) \psi(x,t)=\I \hbar\, 
\frac{\partial\psi(x,t)}{\partial{t}},
\label{9.7}\end{equation}
where the potential $V$ is given by
\begin{equation}
V(x,t)=-e\varphi(x,t){I}_4+e\sum_{n=1}^{3}{\alpha}_{n}A_{n}(x,t).                  \label{9.8}\end{equation}

\paragraph{2. Dirac equation, momentum representation.}
 Further we use the natural units measure, that is, $c=\hbar=1$. 
In the momentum space,  Dirac equation  takes the form
\begin{equation}H(q)\Phi(t,q)+{\int}_{\BR^3}U(q-r)\Phi(t,r)dr=\I\,\frac{\partial\Phi(t,q)}{\partial{t}},
\label{9.10}\end{equation}
where $x=(x_1,x_2,x_3)$,\, $q=(q_1,q_2,q_3)$,\, $r=(r_1,r_2,r_3)$, 
\begin{align}& H(q)=\left[ \begin{array}{cccc}
                          m & 0 & q_3 & q_1-\I q_2 \\
                          0 & m & q_1+\I q_2  & -q_3 \\
                          q_3 & q_1-\I q_2  & -m & 0 \\
                          q_1+\I q_2  & -q_3 & 0 & -m
                        \end{array}\right] .
\label{9.9}
\\ &
\Phi(t,q)={\int}_{\BR^3}\E^{\I qx}\psi(x,t)dx,\quad U(q)={\int}_{\BR^3}\E^{\I qx}V(x)dx.
\label{9.11}\end{align}
The eigenvalues $\lambda_{k}$ and the corresponding eigenvectors
$g_k$ of  $H(q)$ are important, and we find them below:
\begin{equation}
\lambda_{1,2}=-\sqrt{m^2+|q|^{2}},\quad \lambda_{3,4}=\sqrt{m^2+|q|^{2}}
\quad ( |q|^2:=q_{1}^2+q_{2}^2+q_{3}^2);
\label{9.13}\end{equation}
\begin{equation}
g_1=\begin{bmatrix}(-q_1+\I q_2)/(m+\lambda_3) \\ q_3/(m+\lambda_3) \\ 0 \\ 1\end{bmatrix}, \quad
g_2=\begin{bmatrix}-q_3/(m+\lambda_3) \\ (-q_1-\I q_2)/(m+\lambda_3) \\1 \\ 0 \end{bmatrix},
\label{9.14}
\end{equation}
\begin{equation}g_3=\begin{bmatrix} (-q_1+\I q_2)/(m-\lambda_3)\\ q_3/(m-\lambda_3)\\ 0 \\ 1\end{bmatrix},
\quad
g_4= \begin{bmatrix} -q_3/(m-\lambda_3) \\ (-q_1-\I q_2)/(m-\lambda_3) \\ 1 \\ 0\end{bmatrix}.
\label{9.17}\end{equation}

We need the linear spans of some sets $\{g_k\}$ of  vectors $g_k$ and we introduce the following
notations:
\begin{align}
& \label{9!}
M_1(q)= {\mathrm{Span}} \{g_k(q): \, k=1,2\}, \quad M_2(q)= {\mathrm{Span}} \{g_k(q): \, k=3,4\}.
\end{align}
We shall use the fact that  the eigenvectors $g_{1}(q)$ and $g_{2}(q)$ of $H(q)$ have the same eigenvalues
$\lambda_{1,2}=-\sqrt{m^2+|q|^{2}}$, and the eigenvectors $g_{3}(q)$ and $g_{4}(q)$  have the same eigenvalues  $\lambda_{3,4}=\sqrt{m^2+|q|^{2}}$.
\paragraph{3.}
 Let us introduce the self-adjoint operators
\begin{equation}
A_{0}\Phi(q)=H(q)\Phi(q),\quad A\Phi(q)=H(q)\Phi(q)+{\int}_{\BR^3}U(q,r)\Phi(r)dr.
\label{5.17}\end{equation}
Equations \eqref{9.10} form a subclass of  the class of equations
\begin{equation}
\I\frac{\partial\Phi(t,q)}{\partial{t}}=A\Phi(t,q).
\label{5.18}\end{equation}
\begin{Tm}\label{Theorem 5.1}Let the generalized wave operators $W_{\pm}(A,A_0)$
exist. Then the corresponding scattering operator $S(q)$ and the deviation factor
$\wt W_{0}(t,q)$ have the following structure:
\begin{equation}\label{5.19}
S(q)=\sum_{n=1}^{2}s_{n}(q), \qquad
\wt W_{0}(t,q)=\sum_{n=1}^{2}w_{n}(t,q), 
\end{equation}
where  
\begin{equation}\nonumber
s_{n}(q)=P_{n}(q)S(q)P_{n}(q), \qquad 
w_{n}(t,q)=P_{n}(q)\wt W_0(t,q)P_{n}(q) , 
\end{equation}
$P_{n}(q)$
is the orthogonal projector of the space $\BC^4$ onto $M_{n}(q)$,
and $M_n(q)$ $(n=1,2)$ are introduced  in \eqref{9!}.
\end{Tm}
We note that $S(q)$ and $\wt W(t,q)$ are unitary transformations. Hence, the transformations
$s_{n}(q)$ and $w_{n}(t,q)$     are unitary transformations in the spaces $M_{n}(q)$ $(n=1,2)$.

\section{Scattering operator in quantum \\ electrodynamics}\label{S7}
\setcounter{equation}{0}
\paragraph{1.}
The fundamental equation of quantum electrodynamics (often called  \emph{interaction picture} \cite[p. 273]{AB}, see also \cite{Sred}) has the form
\begin{equation}
\I\frac{\partial\Phi(t)}{\partial{t}}={\varepsilon} \clv (t)\Phi(t), \label{6.1}
\end{equation}
where $\Phi(t)$ is  the wave operator function which describes
the state of the field at the time $t$ and $\varepsilon$ is the small parameter.
Rewrite \eqref{6.1} in the following way:
\begin{equation}\Phi(t)=S(t,t_0)\Phi(t_0); \quad \I\, \frac{\partial{S(t,t_0)}}{\partial{t}}={\varepsilon} \clv (t)S(t,t_0),\quad
S(t_0,t_0)=I.\label{6.2}\end{equation}
Now,  we consider the series expansion of $S(t,t_0)$ and write down the recursive formulas for coefficients:
\begin{align}&
S(t,t_0)=\sum_{k=0}^{\infty}{\varepsilon}^{k}S_{k}(t,t_0); \quad S_{0}(t,t_0)=I,\label{6.4}
\\ &
S_{k}(t,t_0)=-\I {\varepsilon}\int_{t_0}^{t} \clv (u)S_{k-1}(u,t_0)du,\quad k>0.
 \label{6.5}\end{align}

\paragraph{2.}
 Let $\Phi(-\infty)$ and $\Phi(+\infty)$ be the operators which describe
the states of the field at the time $t=-\infty$ and $t=+\infty$, respectively.
In view of \eqref{6.2} we have
\begin{equation}\Phi(+\infty)=S(+\infty,-\infty)\Phi(-\infty),
\label{6.6}\end{equation}
where $S(+\infty,-\infty)$ is the scattering operator. It is often assumed that the initial and final states of the system are free,
that is,  $ \clv (\pm{\infty})=0.$
However, in many important cases the initial and final states are not free, and  we shall show that in these cases the theory of generalized wave and generalized scattering operators 
is useful.  In particular, the deviation factors
$W_{-}(t)$ and $W_{+}(t)$ describe the deviation of the initial and final states from
the free state.

\paragraph{3.}
Dirac operators in the presence of the electromagnetic fields (and in momentum representations)
have the form \cite[Ch. IV]{AB}:
\begin{equation}A_{0}\Phi(q)={\wh H}(q)\Phi(q), \quad
 A\Phi(q)={\wh H}(q)\Phi(q)+{\int}U(q,r)\Phi(r)dr,\label{6.7}\end{equation}
 where
\begin{equation}{\wh H}(q)=\left[\begin{array}{cc}
                                     H(q) & 0 \\
                                     0& H(q)
                                   \end{array}\right],\label{6.8}\end{equation}
and  $H(q)$ is given by  \eqref{9.9}.  The fundamental equation (Schr\"odinger picture) in the momentum space has the form
\begin{equation}
\I\,\frac{\partial\Phi(q,t)}{\partial{t}}=A\Phi(q,t).
\label{6.11}\end{equation}
Further we consider the scattering matrix $S(+\infty,-\infty)$ in the momentum representation.
In order to consider $S(q)$ we recall the results on eigenvectors of $H(q)$ in Section \ref{S6}
and easily write down the eigenvectors $G_k\in \BC^8$ of   ${\wh H}(q)$:
\begin{equation}G_{k}=\begin{bmatrix}g_{k} \\ 0\end{bmatrix} \,\, {\mathrm{for}} \,\, 1\leq{k}\leq{4}; \quad
G_{k}=\begin{bmatrix}0 \\ g_{k-4} \end{bmatrix} \,\, {\mathrm{for}} \,\, 5\leq{k}\leq{8}.
\label{6.9}\end{equation}
where   $g_{k}\in \BC^4$ are given by  \eqref{9.14} and \eqref{9.17}. We set
\begin{align}\label{N} &
N_1(q)= {\mathrm{Span}} \{G_k: \, k=1,2,5,6\},
 \\           \label{N1} &
N_2(q)= {\mathrm{Span}} \{G_k: \, k=3,4,7,8\}.
\end{align}
Clearly, the eigenvectors $G_{1}(q),\, G_{2}(q),\,G_{5}(q)$ and $G_{6}(q)$ have equal eigenvalues
$\lambda_{1,2,5,6}=-\sqrt{m^2+|q|^{2}}$, and the eigenvectors $G_{3}(q),\, G_{4}(q),\,G_{7}(q)$ and $G_{8}(q)$   have equal eigenvalues  $\lambda_{3,4,7,8}=\sqrt{m^2+|q|^{2}}$.

\begin{Tm}\label{Theorem 6.1}
Let the generalized wave operators $W_{\pm}(A,A_0)$, where $A$ and $A_0$ are given by \eqref{6.7},
exist. Then the corresponding scattering operator $S(q)$ and the deviation factor
$\wt W_{0}(t,q)$ have the following structure$:$
\begin{align}&
S(q) =s_{1}(q)+s_{2}(q), \label{5.12}\\ &
\wt W_{0}(t,q) = w_{1}(t,q)+w_{2}(t,q), \label{5.13}
\end{align}
where   
$$s_{n}(q)=\wh{P}_{n}(q)S(q)\wh{P}_{n}(q), \quad  w_{n}(t,q)=\wh{P}_{n}(q)\wt W_{0}(t,q)\wh{P}_{n}(q),$$   
$\wh{P}_{n}(q)$
is the orthogonal projector of the space
$\BC^8 $
onto  $N_{n}(q)$, and  $N_n(q)$ $(n=1,2)$ are introduced in \eqref{N} and \eqref{N1}.
\end{Tm}
We note that $S(q)$ and $\wt W(t,q)$ are unitary transformations. Hence, the transformations
$s_{n}(q)$ and $w_{n}(t,q)$   are unitary transformations in the spaces  $N_{n}(q)$.
Further we shall use the next corollary.
\begin{Cy}\label{Example 6.2} Assume that there are such $k$ and $\ell$ $(1 \leq k,\ell \leq 8)$
that all the entries of the $k$-th row and of the $\ell$-th column of the generalized scattering matrix $S(q)$
$($excluding  $s_{k\ell}(q))$ equal zero.
Then $s_{k \ell}(q)$ satisfies
the equality
\begin{equation} |s_{k \ell}(q)|=1.\label{5.15}\end{equation}
\end{Cy}
\section{Example, logarithmic type singularity}\label{S8}
\setcounter{equation}{0}
Consider the series expansion of the scattering function:
\begin{equation}{S}(t,\tau,q)=\sum_{k=0}^{\infty}{\varepsilon}^{k}{S}_{k}(t,\tau,q).
\label{S1}\end{equation}
The following conditions are fulfilled for a number of problems in quantum
 electrodynamics (see \cite{AB} and \cite{BLP}):
\begin{enumerate}[1) ]
	\item The  function $S_{1}(t,\tau,q)$ is uniformly bounded when both inequalities
	$t>1$ and $\tau<-1$ hold.
	\item The function $S_{2}(t,\tau,q)$ has the form
\begin{equation}S_{2}(t,\tau,q)={\varepsilon}^{2}\Big( \varphi(q)\ln{|t\tau|}+O(1)\Big) ,\quad t{\to}\infty,\quad \tau{\to}-\infty .
\label{8.1}\end{equation}
\end{enumerate} 
In other words, the function $S_{2}(t,\tau,q)$ has a logarithmic type singularity. 

Let the conditions 1) and 2) above be fulfilled, and introduce the functions
\begin{equation}
\wt W_{+}(t,q)=t^{\I {\varepsilon}^{2}\varphi(q)}\,\, (t>0),\quad 
\wt W_{-}(\tau,q)=|\tau|^{-\I {\varepsilon}^{2}\varphi(q)}\,\,(\tau<0),\label{8.2}\end{equation}
Similar to \eqref{S1}, the generalized scattering function
\begin{equation}\widetilde{S}(t,\tau,q)=\wt W_{+}(t,q)^*S(t,\tau,q)\wt W_{-}(\tau,q)
\label{8.3}\end{equation}
admits a series expansion: 
\begin{equation}\widetilde{S}(t,\tau,q)=\sum_{k=0}^{\infty}{\varepsilon}^{k}\widetilde{S}_{k}(t,\tau,q).
\label{8.4}\end{equation}
It is easy to see that the following assertion is valid.

\begin{Pn}\label{Proposition 8.1}
The  function $\widetilde{S}_{2}(t,\tau,q)$ is uniformly bounded when both inequalities $t>1$ and $\tau<-1$ hold.
\end{Pn}
\begin{Rk} Earlier we proved the absence of divergences in some important cases of  Schr\"odinger and Dirac equations with Coulomb
potentials $($see \cite{Sakh2, Sakh12} as well as Remark \ref{Remark 2.1} here$)$. Moreover, the existence of the generalized wave operators
implies the existence of the generalized scattering operator and absence of divergences in the cases considered in Proposition \ref{Proposition 3.3}
and in Theorem \ref{Theorem 6.1}. 
\end{Rk}
Below we consider one of the examples where the coefficient $S_2(t,\tau, q)$ (of the non-generalized scattering function) has a logarithmic type singularity and 
the condition \eqref{8.1} is fulfilled.
\begin{Ee}\label{Example 8.2}
Introduce  operators
\begin{align} &
\mathcal{L}_{0}f=x^{2}f,\quad f{\in}L^{2}(0,\infty),\label{8.5}
\\ & 
 \mathcal{L}f=x^{2}f+\int_{0}^{\infty}f(y)R(x,y)dy,\label{8.6}
\end{align}
where
\begin{align}&
R(x,y)=\varepsilon^{2}p(x)p(y)\ln{|x^2-y^2|},\quad p(x)>0.
\label{8.7}\end{align}
\end{Ee}
\begin{Pn}\label{Proposition 8.3}
Let $p(x)$ satisfy the condition
\begin{equation}|p(x_2)-p(x_1)|{\leq}C|x_2-x_1|^{\alpha},\quad \alpha>0.\label{8.8}\end{equation}
Then the equality \eqref{8.1} is valid for $S_2(t,\tau, q)$ corresponding to the operators
$\mathcal{L}_{0}$ and $\mathcal{L}$ above and for
$\varphi(q)=- \pi p^{2}(q)/(2q)$.
\end{Pn}

\begin{proof} 
It is easy to see that
\begin{equation}S_{2}(t,\tau)f=
-\I \int_{\tau}^{t}\int_{0}^{\infty}f(y)R(x,y)
\E^{\I (x^{2}-y^{2})t_{1}}dydt_{1}.\label{12.9}\end{equation}
Let the function $f(x)$ be differentiable  and such that $f(x)=0$ for
all $x \, {\notin} \, [0, \, M]$ (with  some fixed $M>0$). 
Introduce new variables $v=x^2$ and $u=y^2$.
Using \eqref{8.8} and \eqref{12.9} we obtain
\begin{equation}S_{2}(t,\tau)f{\sim}
-f(x)\frac{p^{2}(x)}{2x}\int_{0}^{\sqrt{M}}\big(\ln{|u-v|}\big)\frac
{\E^{\I (u-v)t}-\E^{\I (u-v)\tau}}{u-v}dv.\label{8.10}\end{equation}
We have
\begin{equation}\int_{0}^{\sqrt{M}}\big(\ln{|u-v|}\big)\frac
{\E^{\I (u-v)t}-\E^{\I (u-v)\tau}}{u-v}dv=\int_{u-\sqrt{M}}^{u}(\ln{| \xi |})\frac
{\E^{\I  \xi t}-\E^{\I  \xi\tau}}{ \xi}d \xi.\label{8.11}\end{equation}
The assertion of the proposition follows from \eqref{8.10}, \eqref{8.11}
and the equality:
\begin{equation}\int_{0}^{\infty}\frac{\sin{ \xi}}{ \xi}d \xi=\pi/2.\label{8.12}\end{equation}
\end{proof}
\section{Series expansions of the entries \\ of scattering matrices}\label{Exp}
In cases of various diagrams, there is a series expansion
\setcounter{equation}{0}
\begin{equation}s_{k \ell}(q,L, \ve)=1+{\ve}a_{1}(q)+{\ve}^{2}a_{2}(q,L)+...,
\label{14.4}\end{equation}
where
\begin{equation}
a_{2}(q,L)=\int_{\Omega}F(p,q)d^{4}p,\label{14.2}\end{equation}
$\Omega$ is a four dimensional sphere (invariant region of integration) with radius $L$,
$p=[- \I p_0,p_1,p_2,p_3],$ and $F$ is a rational function with respect to $p$ and $q$ (see, e.g., \cite[p. 631]{AB} and \cite{Fri}).
We shall consider the cases where the limit on the right-hand side of \eqref{14.2}
does not exist when $L$ tends to infinity.
\begin{Ee}\label{Example 14.1} Let the relation \begin{equation}a_{2}(q,L)=\int_{\Omega}F(p,q)d^{4}p=\I \big(\phi(q)\ln{L}+\psi(q)+o(1)\big)\label{14.3}\end{equation}
be valid for some real valued functions
$\phi(q)$ and ${\psi(q)}$.
Then, the  integral in  \eqref{14.3} diverges logarithmically, and the second term 
\begin{align} & a_{2}(q):=\lim_{L \to \infty}\int_{\Omega}F(p,q)d^{4}p \label{E}
\end{align}
in the power series 
\begin{align}& s_{k \ell}(q, \ve)=1+{\ve}a_{1}(q)+{\ve}^{2}a_{2}(q)+...,
\label{E1}
\end{align}
which corresponds to the entry $s_{kl}$ of the non-generalized scattering function,
is equal to infinity.
\end{Ee}
In order to remove this divergence,
we introduce
\begin{equation}\widetilde{s}_{k \ell}(q,L, \ve)=
L^{-\I {\varepsilon}^{2}\phi(q)}{s}_{k \ell}(q,L, \ve).
\label{14.6}\end{equation}
Using \eqref{14.4} and \eqref{14.6} we have
\begin{equation}\widetilde{s}_{k \ell}(q,L, \ve)=1+{\varepsilon}a_{1}(q)+{\varepsilon}^{2}
\big(a_{2}(q,L)-\I \phi(q)\ln{L}\big)+...
\label{14.7}\end{equation}
It follows from \eqref{14.3}  that the second term
\begin{equation}\widetilde{a}_{2}(q,L):=a_{2}(q,L)-\I \phi(q)\ln{L} \label{14.8}\end{equation}
of the power series \eqref{14.7} converges when $L{\to}\infty$.

We emphasize  that
\begin{equation}|{s}_{k \ell}(q,L, \ve )|=|\widetilde{s}_{k \ell}(q,L,\ve)|.\label{14.9}\end{equation}
\begin{Rk}\label{Remark 14.2}
The factor $U_{0}(q, L, \ve)=L^{\I {\varepsilon}^{2}\phi(q)}$ is an analogue of the deviation factor $\wt W_{0}(t,q)$ 
which we considered in the previous sections. Namely, the  factor $U_{0}(q,L, \ve)$
describes the deviation of the initial and final states of a system from
the free state.\end{Rk}
\begin{Rk}\label{Remark 14.3} It is  well known  that 
 condition \eqref{14.3} is fulfilled for many problems arising in the theory of  collisions of particles.\end{Rk}

\begin{Ee}\label{Example 14.4} Let the relation \begin{equation}a_{2}(q,L)=\int_{\Omega}F(p,q)d^{4}p=\I\big(\phi(q)L^{2}+\psi(q)L+{\nu}(q)\ln{L}
+\mu(q)+o(1)\big),\label{14.10}\end{equation}
where $\phi, \psi, \nu$ and $\mu$ are real valued functions, be valid.
In this $($more general$)$ case, the factor  $U_{0}(q,L,\ve)$ has the form
\begin{equation} U_{0}(q,L,\ve)=\E^{\I{\varepsilon}^{2}\big(\phi(q)L^{2}+\psi(q)L\big)}L^{\I{\varepsilon}^{2}\nu(q)}.
\label{14.11}\end{equation} Using $U_0$, we again introduce $\widetilde{s}_{k \ell}:$
\begin{equation}\widetilde{s}_{k \ell}(q,L,\ve)=
U_{0}(q, L,\ve)^{-1}{s}_{k \ell}(q,L, \ve).
\label{14.13}\end{equation}
Now, in view of \eqref{14.10},
relations \eqref{14.7} and \eqref{14.8}  take the forms$:$
\begin{align}& \widetilde{s}_{k \ell}(q,L, \ve)=1+{\varepsilon}a_{1}(q)+{\varepsilon}^{2}
\widetilde{a}_{2}(q,L)+...
\label{14.14}\\ &
\widetilde{a}_{2}(q,L)=a_{2}(q,L)-\I \big(\phi(q)L^{2}+\psi(q)L+{\nu}(q)\ln{L}\big) \label{4.15!}\end{align}
It follows from \eqref{14.10} and \eqref{4.15!} that the second term in the power series \eqref{14.14} converges when $L{\to}\infty$.
\end{Ee}
Relation \eqref{14.9} also holds  for Example \ref{Example 14.4}.
\begin{Rk}\label{Remark 14.5} In the cases of all irreducible diagrams, the relation \eqref{14.10} is valid $($see \cite[Sect. 46, 47]{AB}$)$.
\end{Rk}
The simplest subcase of Example \ref{Example 14.4} is obtained by setting 
\begin{equation}\phi(q)=0,\quad \nu(q)=0,\quad \psi(q)=1.\label{14.16}\end{equation}
In this subcase we have
\begin{equation}U_{0}(q, L, \ve)=U_{0}(L, \ve)=\E^{\I \varepsilon^{2}L}.\label{14.17}\end{equation}

\section{Conclusion}
\setcounter{equation}{0}
We have shown that generalized scattering operators have an interesting inner structure 
(see Theorems \ref{Theorem 5.1} and \ref{Theorem 6.1}). We believe that this structure
(e.g., relation \eqref{5.12}) may be checked experimentally.

In Remark \ref{Remark 2.1}
and in Sections \ref{S8} and \ref{Exp} we discuss the absence of divergences when
one uses generalized scattering operators instead of the classical scattering operators.
Thus, generalized scattering operators present an important tool to avoid
divergences.  Moreover, generalized scattering operators exist for the cases
where the initial and final states of the system are not free, and the classical scattering
operators do not exist.

\vspace{0.3em}
The dynamical and stationary deviation factors $W_0(t)$
and $V_0(r)$  describe the deviation of the initial and final states of  the corresponding
system from the free state. This fact can be checked , I think, by experimental way
Deviation factors are not uniquely defined. The non-uniqueness
is contained in constant operator multipliers (multipliers $C_{\pm}$ for the case of $W_0(t)$ 
and multiplier $C$ for $V_0(r)$). The choice of multipliers $C_{\pm}$  and $C$ depends on the particular physical problem
under consideration. 

The following principles of choosing the constant multipliers $C_{\pm}$ and $C$
can be formulated for radial Schr\"odinger and Dirac equations.
In the case of radial Schr\"odinger  equations (considered in Sections \ref{Coul} and \ref{Erg}),
the  operators $C_{\pm}$ and $C$ should not depend on parameter $\ell$.
In the case of radial Dirac  equation (considered in Section \ref{radDir}),
the  operators $C_{\pm}$ and $C$ should not depend on parameter $k$.
{\it Then, the effective cross section of the scattering does not depend on
these multipliers.}

Important ergodic interrelations between dynamical and stationary deviation factors are
given in formulas \eqref{3.21}, \eqref{d3} and \eqref{11.43}. We suppose that these
interrelations may be confirmed experimentally and that similar interrelations hold in other important cases.

\vspace{0.3em}

Recall that  $S_{st}$ and $S_{dyn}$ stand for the generalized stationary and dynamical, respectively,  scattering operators.

{\bf Open problem.}
Prove the equality 
\begin{equation}S_{st}=S_{dyn}.\label{12.1}\end{equation}
We  proved \cite{Sakh2, Sakh12} the  equality \eqref{12.1} for the cases
of Schr{\"o}dinger and Dirac equations with Coulomb potentials.

\vspace{0.3em}

{\bf Acknowledgements.}
The author is grateful to  A. Sakhnovich and I.~Roitberg for fruitful discussions
and help in the preparation of the manuscript, and to I. Tydniouk for his help with
some important calculations in Section~\ref{S6}.


\begin{thebibliography}{11}  
\bibitem{AB}
 A.I. Akhiezer and  V.B. Berestetskii, ``\textit{Elements of Quantum Electrodynamics}" (Davey, New York, 1964).

\bibitem{AG}
 N.I. Akhiezer and I.M. Glazman, ``\textit{Theorie der linearen Operatoren im Hilbert-Raum}" (Akademie-Verlag, Berlin, 1958).

\bibitem{AlK}
D. Alpay and B. Kirstein (eds), ``\textit{Recent advances in inverse scattering, Schur analysis and stochastic processes. A collection of papers dedicated to Lev Sakhnovich}", 
  Operator Theory Adv. Appl., Vol. 244: Linear Operators and Linear Systems (Birkh\"auser/Springer, Cham, 2015).   


	
\bibitem{BEKT}
A. Beigl, J. Eckhardt, A. Kostenko, and G. Teschl, 
``On Spectral Deformations and Singular Weyl Functions for One-Dimensional Dirac Operators", 
J. Math. Phys. \textbf{56}, 012102 (2015).    

\bibitem{Bel}
R. Bellman, ``\textit{Stability theory of differential equations}" (McGraw-Hill Book Co., New York, 1953).   

\bibitem{BLP}
V.B. Berestetskii, E.M. Lifshitz, and L.P. Pitaevskii, ``\textit{Quantum electrodynamics}" (Butterworth-Heinemann, 1982).   

\bibitem{Ber}
F.A. Berezin, ``On a model for quantum field theory",  Math. USSR Sbornik \textbf{5}:1, 1--23 (1968).   

\bibitem{BetSal}
H.A.  Bethe and E.E. Salpeter, ``\textit{Quantum mechanics of one- and two-electron atoms}" (Springer, Berlin, 1957).
  
  
\bibitem{Bur}
P.G. Burke, ``\textit{R-matrix theory of atomic collisions. Application to atomic, molecular and optical processes}" (Springer, Berlin, 2011).  
 
\bibitem{BuM}
V.S. Buslaev and  V.B. Matveev, ``Wave operators for the Shr{\"o}dinger equation with a slowly
decreasing potential", Theor. Math. Phys. \textbf{2}:3, 266--274 (1970).   

   
\bibitem{Do}
J. Dollard, ``Asymptotic convergence and Coulomb interaction", J. Math. Phys. \textbf{5}, 729--738 (1964).

\bibitem{EH}
H. Elvang and Y.-T. Huang, ``\textit{Scattering amplitudes in gauge theory and gravity}" (Cambridge University Press, Cambridge, 2015).

\bibitem{FaMe}
 L.D. Faddeev and S.P. Merkuriev,  ``\textit{Quantum scattering theory for several particle systems}" (translated from the 1985 Russian original),  
 Mathematical Physics and Applied Mathematics, Vol. 11 (Kluwer, Dordrecht, 1993).

  
\bibitem{Fri}
H.M. Fried, ``\textit{Modern functional quantum field theory. Summing Feynman graphs}" (World Scientific, Hackensack, NJ, 2014).    

\bibitem{GiVe}
 J. Ginibre and  G. Velo,  ``Modified wave operators without loss of regularity for some long range Hartree equations", II,  Comm. Pure Appl. Anal. \textbf{14}:4, 1357--1376 (2015).




\bibitem{JMG}
J.M. Jauch, B. Misra, and A.G. Gibson, ``On the asymptotic condition of scattering theory", Helv. Phys. Acta \textbf{41}:4, 513--527 (1968). 
  
\bibitem{Kat}
T. Kato, ``Perturbation of continuous spectra by trace class operators", 
Proc. Japan Acad. \textbf{33}:5, 260--264 (1957).   

\bibitem{KoSaTe}
A.~Kostenko, A.~Sakhnovich, and G.~Teschl, ``Weyl--Titchmarsh theory for Schr\"odinger operators with strongly singular potentials", 
Int. Math. Res. Not. \textbf{2012}:8, 1699--1747 (2012).  
 
\bibitem{KuKu}
H. Kubo and  K. Kubota, ``Generalized wave operators for a system of semilinear wave equations in three space dimensions", Hokkaido Math. J. \textbf{42}:1, 81--111 (2013).  

   

   
\bibitem{Mat}
V.B.  Matveev, ``Invariance principle for generalized wave operators",  Theor. Math. Phys. \textbf{8}:1, 663--667 (1971).

\bibitem{Nak}
S. Nakamura, ``Modified wave operators for discrete Schr\"odinger operators with long-range perturbations", 
J. Math. Phys. \textbf{55}:11, 112101 (2014). 

\bibitem{OC} 
 M. O'Carroll, ``The existence and completeness of generalized wave operators for spherically symmetric, asymptotically Coulomb potentials",  J. Math. Phys. \textbf{13}, 268--271 (1972).

 \bibitem{Pear}
 D.B. Pearson,  ``Conditions for the existence of the generalized wave operators", J. Math. Phys. \textbf{13}, 1490--1499 (1972).

\bibitem{Ros}
M. Rosenblum, ``Perturbation of continuous spectrum and unitary equivalence", 
Pacif. J. Math. \textbf{7}:1, 997--1010 (1957).   

\bibitem{SaSaR}
A.L. Sakhnovich,  L.A. Sakhnovich, and I.Ya. Roitberg,   ``\textit{Inverse problems and nonlinear evolution equations. 
 Solutions, Darboux matrices and Weyl-Titchmarsh functions}", {De Gruyter Studies in Mathematics, Vol. 47} (De Gruyter, Berlin, 2013).
 
\bibitem{Sakh8}
L.A. Sakhnovich, ``Dissipative operators   with absolutely continuous spectrum", 
Trans. Moscow Math. Soc. \textbf{19}, 233--297 (1968).   
    
\bibitem{Sakh2}
L.A. Sakhnovich, ``Generalized wave operators", Math. USSR Sbornik \textbf{10}:2,  197--216 (1970).   

\bibitem{Sakh12}
L.A. Sakhnovich, ``The invariance principle for generalized wave operators", 
Functional Anal. Appl. \textbf{5}:1, 49--55 (1971).   


\bibitem{Sakh11}
L.A. Sakhnovich, ``On properties of discrete and continuous spectra of
Dirac radial equation", Theor. Math. Phys. \textbf{108}:1, 876--888 (1996).   


 

 
   
\bibitem{Sred}
 M. Srednicki, ``\textit{Quantum Field Theory}"  (Cambridge University Press,  Cambridge, 2007).    

\bibitem{Str}
R.F.  Streater, ``\textit{A theory of scattering for quasifree particles}"  
(World Scientific, Hackensack, NJ, 2015).


 
\bibitem{Tong}
 Y.S. Tong,  ``Generalized wave operators in space with an indefinite metric", J. Math. Anal. Appl. \textbf{137}:2, 371--395 (1989).


\bibitem{Wir}
J. Wirth, ``Scattering and modified scattering for abstract wave equations with time-dependent dissipation", 
Adv. Differential Equations \textbf{12} (2007), 1115--1133.

\bibitem{Xi}
J. Xia, 
``Spectral convergence of selfadjoint operators and generalized wave operators",  J. Funct. Anal. \textbf{77}:1 176--197 (1988).

\end{thebibliography}
\end{document}